\documentclass[aps,twocolumn]{revtex4}
\usepackage{graphicx}
\usepackage{dcolumn}
\usepackage{amsmath}
\usepackage[latin1]{inputenc}

\begin{document}

\title[Short Title]{Multi-excitonic complexes in single InGaN quantum dots}

\author{R. Seguin\footnote{email: seguin@sol.physik.tu-berlin.de}, S. Rodt, A. Strittmatter, L. Rei{\ss}mann, T. Bartel, A. Hoffmann, and D. Bimberg}
\affiliation{Institut f\"ur Festk\"orperphysik, Technische
Universit\"at Berlin, D-10623 Berlin, Germany}

\author{E. Hahn and D. Gerthsen}
\affiliation{Laboratorium f\"ur Elektronenmikroskopie,
Universit\"at Karlsruhe, D-76128 Karlsruhe, Germany}

\begin{abstract}
Cathodoluminescence spectra employing a shadow mask technique of
InGaN layers grown by metal organic chemical vapor deposition on
Si(111) substrates are reported. Sharp lines originating from
InGaN quantum dots are observed. Temperature dependent
measurements reveal thermally induced carrier redistribution
between the quantum dots. Spectral diffusion is observed and was
used as a tool to correlate up to three lines that originate from
the same quantum dot. Variation of excitation density leads to
identification of exciton and biexciton. Binding and anti-binding
complexes are discovered.
\end{abstract}

\maketitle

The fundamental processes responsible for optical recombination in
InGaN/GaN quantum structures are a matter of controversial
discussion. Wurzite GaN-based semiconductors exhibit strong
piezoelectric fields owing to their large piezoelectric constants.
In thin layers these fields give rise to a blueshift of the
transition energy with increased excitation density due to
screening of the quantum confined Stark effect.$^1$ Yet, the quantum dot (QD) nature of
compositional fluctuations in the InGaN layers can lead to
similar effects.$^{2,3}$ The purpose of
the present work is to demonstrate three-dimensional confinement of carriers in
InGaN by conducting spatially high-resolved cathodoluminescence
(CL) measurements. We observe a multitude of sharp lines, several
of these lines unambiguously originating from the same QD.
Temperature and excitation density dependent CL investigations of
these lines give further proof for the existence of strong
localization and lead to the distinction of excitonic and
biexcitonic recombination.

The samples were grown by low-pressure metal organic chemical vapor deposition using a horizontal
AIX200 RF reactor. Prior to loading the reactor the substrates
were treated by wet chemical etching yielding an oxide-free, and
hydrogen terminated Si(111) surface. An AlN layer acting as
nucleation surface was obtained by using a previously described
conversion process of AlAs to AlN.$^4$ In the
following step Al$_{0.05}$Ga$_{0.95}$N/GaN buffer layers were
grown at T=1150~$^{\circ}$C up to a total thickness of 1 $\mu$m.
The InGaN layers were grown at 800 $^{\circ}$C using TMGa, TMIn,
and ammonia as precursors. Total pressure was kept at 400 mbar
during InGaN deposition. The growth was finished with a 20 nm GaN
cap layer grown during heatup to 1100 $^{\circ}$C.

The samples were investigated with a JEOL JSM 840 scanning
electron microscope equipped with a cathodoluminescence setup.$^5$ All measurements were made at a temperature
of 6.5 K unless stated otherwise. The low temperatures were
obtained by mounting the samples onto a He flow cryostat. The
luminescence light was dispersed by a 0.3 m monochromator equipped
with a 2400 lines/mm grating and detected with a nitrogen cooled
Si-CCD camera, providing a spectral resolution of 310 $\mu$eV at 3
eV. In order to increase spatial resolution we applied metal
shadow masks onto the sample surface with aperture diameters of
100 and 200 nm.

The integral CL spectrum shows an intensive peak centered at 2.98
eV with a full width at half maximum (FWHM) of 80 meV which
originates from the InGaN layer.
Weak luminescence of the donor bound GaN exciton can be observed
at 3.46 eV.

\begin{figure}[ht]\centering
\resizebox*{0.95\columnwidth}{!}{\includegraphics[angle=0]{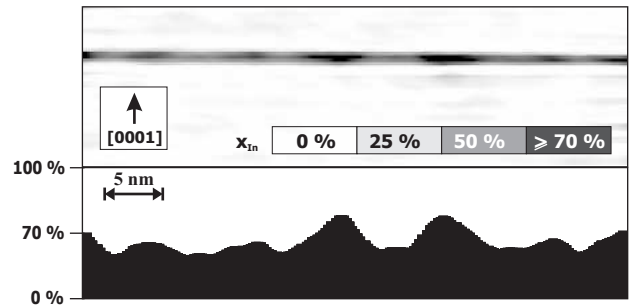}}
\caption{DALI processed cross section HRTEM image. The upper image
shows a gray-scale coded map of the local In concentrations in the
sample. The InGaN layer is about 2~nm in height and shows
prominent In fluctuations. The lower image visualizes the
In concentration along the layer.} \label{TEM Image}
\end{figure}

When measured through one of the apertures the InGaN peak
decomposes into sharp lines. Recently, similar observations on
InGaN structures grown on sapphire  have been reported by other
groups using $\mu$-PL.$^{6,7}$ The narrowest
lines show a FWHM of 0.48 meV. These lines can be found over a
wide range of energies (2.8-3.2 eV) thus covering the whole
ensemble peak. Since the line density is very high we mainly
investigated the high and low energy sides of the ensemble peak
where single lines are well resolved.

DALI$^8$ (digital analysis of lattice images) processed cross-section high-resolution TEM
(HRTEM) measurements, performed under short times of irradiation
($\ll$ 1 min) to prevent electron beam induced artefacts,$^9$ show alternating areas of high and low In
content. The In-rich domains have a lateral size of about 5 nm
(Fig. \ref{TEM Image}). Such fluctuations are small enough to
provide strong localization of carriers and are identified as the
source of the sharp peaks. A similar growth mode of QDs has been
reported for II-VI compounds, such as CdSe/ZnSe. There,
composition fluctuations induce strong carrier localization and
QD-like behavior.$^{10}$

\begin{figure}[ht]\centering
\resizebox*{0.95\columnwidth}{!}{\includegraphics[angle=0]{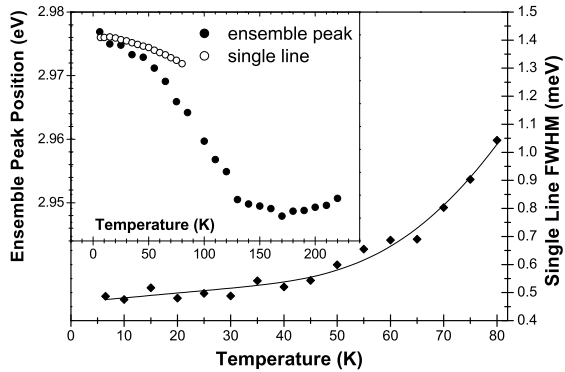}}
\caption{Temperature dependent measurements: FWHM of a single line
with a fit according to Eq.~(\ref{temperatur}). The inset shows
the energetic positions of the ensemble peak and of a single line.
The single line position has an offset for easier comparison.}
\label{temp}
\end{figure}

Temperature dependent measurements of single lines were conducted.
The lines were visible in a temperature range from 6 to 80 K. At
higher temperatures the lines broaden and the peak intensity is
too low to resolve them any more. The energetic position of single
lines shifts about 4 meV to the red with increasing temperature
(inset Fig. \ref{temp}). In the present case (Fig. \ref{temp}) the
FWHM $\Gamma(T)$ increased from initially 0.48 meV, which is
slightly broader than our resolution limit, to about 1.05~meV, at
a rate well below $kT$. $\Gamma(T)$ was fitted well using

\begin{equation} \label{temperatur}
\Gamma (T)= \Gamma_0+\gamma_{p}T+\gamma_{a}
\exp\left(-\frac{E_A}{k_BT}\right).
\end{equation}

$\Gamma_0$ describes the initial FWHM due to the limited spectral
resolution and the spectral diffusion as will be discussed later.
The linear term describes acoustic phonon interaction. The third
term describes dephasing due to excitation of the carriers into
the surrounding InGaN layer. Optical phonons are negligible here
due to their energy of 91.5 meV in GaN, which is much larger than
$kT$ at 6-80 K. The fit yields coupling constants of
$\gamma_{p}=1.7 \pm 0.7$~$\mu$eVK$^{-1}$, $\gamma_a = 36 \pm 25$~
meV and an activation energy of $E_A = 31 \pm 5$~meV. This gives
us an estimate of the localization depth of the probed QD relative
to the surrounding InGaN layer.

Temperature dependent PL measurements of the ensemble peak were
made as well. As can be seen in the inset of Fig. \ref{temp}, the
red shift of the ensemble peak is initially larger than that of a
single line. This behavior is attributed to preferential quenching
of luminescence from small QDs emitting light at the high energy
side of the spectrum. As there are also weakly
localizing QDs in the InGaN layer, the quenching sets in at lower temperatures
compared to, e.g., InAs/GaAs QDs.$^{11}$ Since this is an
ensemble effect it cannot be observed for the single line. At
higher temperatures thermal activation of carriers into higher
energetic states completely compensates the redshift of the
bandgap. This results in an s-shape of the data points as is
characteristic for QD ensembles.

\begin{figure}[ht]\centering
\resizebox*{0.95\columnwidth}{!}{\includegraphics[angle=0]{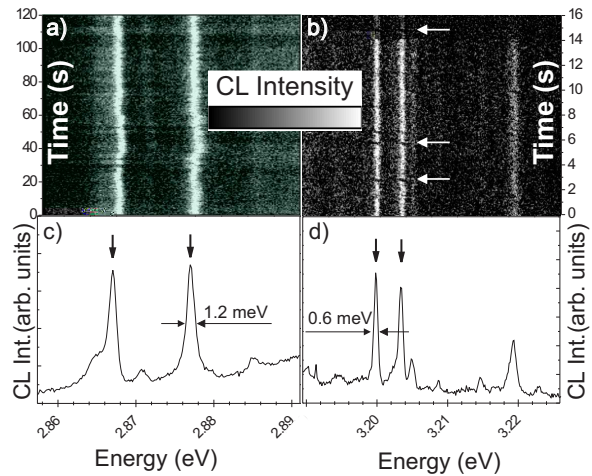}}
\caption{Time series: (a) 400 spectra with 300 ms integration time
each. Spectral diffusion is visible and correlates the marked
lines in (c). The lines are broadened significantly due to the
jitter. (b) 200 spectra with 80 ms integration time each. On/off
blinking (indicated by the white arrows) is visible and correlates
the marked lines in (d). (c) and (d) show averaged spectra of the
time series above.} \label{jitter}
\end{figure}

We took spectra of single lines with short integration time to
generate time series. One series consists of several 100 spectra
with fixed integration times between 80 and 300 ms each. We
observe a slight stochastic variation of the peak energies [Fig.
\ref{jitter}(a)], known as spectral diffusion.$^{12}$ The QDs experience fluctuating electric fields which cause a
slight shift of the energy level in the QD via the quantum
confined Stark effect. The source of these fields is charging and
decharging of nearby defects or interface states. 
When longer integration times are
used, spectral diffusion contributes significantly to the observed
linewidth [Fig. \ref{jitter}(c)] giving rise to a seemingly larger
linewidth than our resolution limit even at low temperatures. At
short integration times, on/off blinking of the lines can be
observed as well [Fig. \ref{jitter}(b)]. The internal electric
fields spatially separate electrons and holes and reduce the
wavefunction overlap or cause the carriers to leave the QD
completely. Strong fields from nearby centers can thus quench the
luminescence completely.$^{13}$

Both effects can be used to correlate transitions that originate
from the same QD since they experience the same electric fields
and hence the lines exhibit the same energetic jitter and blinking
behavior.$^{14}$ We observed doublets and on rare
occasions triplets that show a similar jitter.
Typical energy differences vary between 2 and 20 meV. The presence
of such groups demonstrates the existence of higher excitonic
complexes in one QD.

\begin{figure}[ht]\centering
\resizebox*{0.95\columnwidth}{!}{\includegraphics[angle=0]{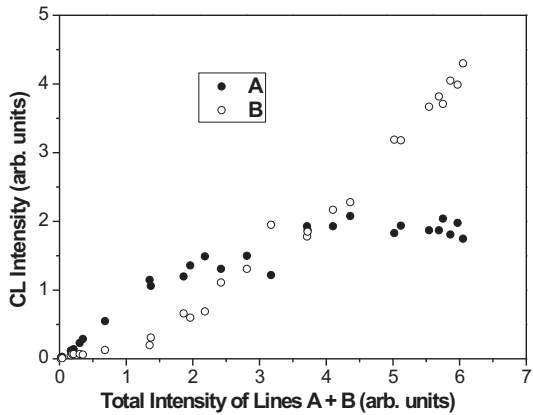}}
\caption{Excitation density dependence. The total intensity equals the
local excitation density. Line A dominates at low excitation
densities and then saturates. Line B takes over at high excitation
densities.} \label{excite}
\end{figure}

In Fig. \ref{excite} the intensity of two lines A and B which had
been correlated by their jitter pattern is plotted versus the
total intensity of both lines. The total intensity is a good
measure for the local excitation density, assuming that the QD
does not saturate and no other lines with similar jitter are
present. Line A depends linearly on the excitation density at low
densities and then saturates. Line B depends quadratically on the
excitation density and dominates in the high excitation regime.
This behavior is typical for exciton and biexciton.$^{15}$ At low excitation densities the probability of
two electron-hole pairs populating the same QD to form a biexciton
before one of them recombines is low. At high excitation densities
the probability increases until it is more likely to find
biexcitons in the QD than excitons. Biexcitons have not yet been
reported for InGaN QD structures before. In the particular case of
Fig. \ref{excite} the biexciton binding energy, which is generally
measured relative to the exciton recombination energy, is -3.5
meV.

B-type lines could be observed at higher and lower energies with
respect to corresponding A-type lines depending on the particular
QD examined. We found values ranging from -3.5 to 16 meV. Hence
binding energies can be both positive and negative. "Anti-binding"
complexes do not exist in bulk semiconductors. In QDs, they can be
stable due to the localizing confinement potential.$^{16}$ Further investigation of the multi-excitonic complexes are under way.

In conclusion, we have demonstrated QD-like behavior of the
luminescence originating from a thin InGaN layer. Composition
fluctuations within the layer identified via HRTEM are the source
for sharp lines in the spectrum. Temperature dependent
measurements  further prove strong localization within the layer.
Observation of spectral diffusion and blinking has been used to
demonstrate the existence of multi-excitonic states. Excitation
density dependent measurements unveiled lines with excitonic and
biexcitonic origin. Binding and anti-binding complexes could be
observed.

Part of this work has been funded by Sonderforschungsbereich 296
of Deutsche Forschungsgemeinschaft.

\noindent $^{1}$ A. Hangleiter, J. Im, H. Kollmer, S. Heppel, J.
Off, and F. Scholz, MRS Internet J. Nitride Semicond. Res.
\textbf{3}, 15 (1998).\\
$^{2}$ S. Chichibu, T. Azuhata, T. Sota, and S. Nakamura, Appl.
Phys. Lett. \textbf{69}, 4188 (1996). \\
$^{3}$ I. L. Krestnikov, N. N. Ledentsov, A. Hoffmann, and D.
Bimberg, Phys. Rev. B \textbf{66}, 155310 (2002). \\
$^{4}$ A. Strittmatter, A. Krost, M. Stra{\ss}burg, V. T\"urck, D.
Bimberg, J. Bl\"asing, and J. Christen, Appl. Phys. Lett.
\textbf{74}, 1242 (1999). \\
$^{5}$ M. Grundmann, J. Christen, N. N. Ledentsov, J. B\"ohrer, D.
Bimberg, S. S. Ruvimov, P. Werner, U. Richter, U. G\"osele, J.
Heydenreich, V. M. Ustinov, A. Yu. Egorov, A. E. Zhukov, P. S.
Kop'ev, and Zh. I. Alferov, Phys. Rev. Lett. \textbf{74}, 4043
(1995) \\
$^{6}$ O. Moriwaki, T. Someya, K. Tachibana, S. Ichida,
and Y. Arakawa, Appl. Phys. Lett. \textbf{76}, 2361 (2000). \\
$^7$ R. A. Oliver, G. A. D. Briggs, M. J. Kappers, C. J.
Humphreys, S. Yasin, J. H. Rice, J. D. Smith, and R. A. Taylor,
Appl. Phys. Lett. \textbf{83}, 755 (2003). \\
$^8$ D. Gerthsen, E. Hahn, B. Neubauer, V. Potin, A. Rosenauer, M.
Schowalter, Phys. Stst. Sol. (c) \textbf{0}, 1668 (2003) \\
$^9$ T. M. Smeeton, M. J. Kappers, J. S. Barnard, M. E. Vickers,
and C. J. Humphreys, Appl. Phys. Lett. \textbf{83}(26), 5419
(2003). \\
$^{10}$ I. L. Krestnikov, M. Stra{\ss}burg, M. Caesar, A.
Hoffmann, U.
W. Pohl, and D. Bimberg, Phys. Rev. B \textbf{60}, 8695 (1999). \\
$^{11}$ R. Heitz, I. Mukhametzhanov, A. Madhukar, A. Hoffmann, and
D. Bimberg, J. Electron. Mater. 28(5), 520 (1999). \\
$^{12}$ S. A. Empedocles, D. J. Norris, and M. G. Bawendi, Phys.
Rev. B \textbf{77}, 3873 (1996).\\
$^{13}$ P. Castrillo, D. Hessman, M.-E. Pistol, and J. A. Prieto,
Jpn. J. Appl. Phys. \textbf{36}, 4188 (1997). \\
$^{14}$ V. T\"urck, S. Rodt, O. Stier, R. Heitz, R. Engelhardt, U.
W. Pohl, and D. Bimberg, Phys. Rev. B \textbf{61}, 9944 (2000).\\
$^{15}$ M. Grundmann and D. Bimberg, Phys. Rev. B \textbf{55},
9740 (1997). \\
$^{16}$ S. Rodt, R. Heitz, A. Schliwa, R. L. Sellin, F. Guffarth,
and D. Bimberg, Phys. Rev. B \textbf{68}, 035331 (2003).
\end{document}